\documentclass[journal=jpccck,manuscript=article]{achemso}

\usepackage{tabularx,booktabs,ragged2e,siunitx}
\usepackage{multirow}
\usepackage{enumitem}

  \usepackage[latin1]{inputenc}

\SectionNumbersOn
\usepackage[version=3]{mhchem} 
\usepackage[sort&compress,numbers,super]{natbib} 
\usepackage{achemso} 
\usepackage{textcomp}
\usepackage{titlesec}
\usepackage{amsmath}
\usepackage{makecell}

\usepackage{fontawesome}

\usepackage{color, colortbl}
\definecolor{Gray}{gray}{0.9}

\author{Luke Melo, Angus Hui, Matt Kowal, Eric Boateng, Zahra Poursorkh, Ed\`ene Rocheron, Jake Wong, Ashton Christy and Edward Grant}

\email{edgrant@chem.ubc.ca}
\affiliation[The University of British Columbia]
{Department of Chemistry, The University of British Columbia \linebreak Vancouver BC, Canada, V6T 1Z1 \linebreak \today}

\title[An \textsf{achemso} demo]
  { Size distributions of gold nanoparticles in solution measured by single-particle mass photometry}

\begin{document}

\pagebreak
 
\begin{abstract}

{Specialized applications of nanoparticles often call for particular, well-characterized particle size distributions in solution.  But, this property can prove difficult to measure.  High-throughput methods, such as dynamic light scattering, detect nanoparticles in solution with an efficiency that scales with diameter to the sixth power.  This diminishes the accuracy of any determination that must span a range of particle sizes.  The accurate classification of broadly distributed systems thus requires very large numbers of measurements.  Mass-filtered particle-sensing techniques offer a better dynamic range, but} are labor-intensive and so have low throughput.  Progress in many areas of nanotechnology requires a faster, lower-cost, and more accurate measure of particle size distributions, particularly for diameters smaller than 20 nm.  {Here, we present a tailored interferometric microscope system, combined with a high-speed image-processing strategy, optimized for real-time particle tracking that determines accurate size distributions in nominal 5, 10, and 15 nm colloidal gold nanoparticle systems by automatically sensing and classifying thousands of single particles sampled from solution at rates as high as 4,000 particles per minute.}  We demonstrate this method by sensing the irreversible binding of gold nanoparticles to poly-D-lysine functionalized coverslips.  Variations in the single-particle signal as a function of time and mass, {calibrated by TEM,} show clear evidence for the presence of diffusion-limited transport that most affects larger particles in solution.

\end{abstract}
\pagebreak

\section*{Introduction}

Colloidal gold nanoparticles have attracted a great deal of attention for their many novel and important properties.  They show a high degree of physical and chemical stability, together with a fundamental biocompatibility that has spurred efforts leading to a host of promising new therapeutic and diagnostic technologies    \cite{boisselier2009gold,dreaden2012golden,li2014anisotropic,zhou2015gold,kong2017unique,elahi2018recent,yu2020size,makvandi2020metal}.  Many cellular and molecular imaging applications exploit the plasmonic characteristics of gold nanoparticles. \cite{murphy2008gold,young2019interferometric,wu2019gold} Related optoelectronic properties also figure in the development of gold nanoparticle biosensors \cite{anker2010biosensing,amendola2017surface}, and sub-wavelength antennae that increase the efficiency of photovoltaic energy conversion \cite{atwater2011plasmonics,linic2021flow}.  Although celebrated for their inertness, conditions exist under which gold nanoparticles can function as high-performance catalysts \cite{haruta1987novel,nkosi1988reactivation,daniel2004gold,hutchings2018heterogeneous,ishida2019importance,du2019atomically}. 

In all of these applications, gold nanoparticles exhibit a suitability for purpose that depends on size.  Virtually all bulk methods of production yield distributions of nanoparticle diameters.  The physical and chemical conditions of synthesis strongly influence the size, shape and coordination of the products in these distributions \cite{jana2001seed,nikoobakht2003preparation,de2019review}.  Purification protocols exist \cite{balasubramanian2010characterization}, but the utility of a separated nanoparticle fraction, particularly for any structure-function determination, depends on an effective means to establish the particle size distribution.  

Existing methods of particle detection, such as dynamic light scattering, can be calibrated to provide indirect information on the size distribution in an aqueous suspension of nanoparticles of adequate size  \cite{balasubramanian2010characterization,rice2013particle,laborda2016detection}.  With suitable mounting, transmission electron microscopy can survey evaporated samples \cite{sweeney2006rapid,zhao2013state}.  Separation methods, such as electrophoresis and hydrodynamic chromatography, often mass-resolve particle fraction distributions by inductively coupled plasma-mass spectrometry  \cite{helfrich2006size,franze2014fast}.  All of these methods require the measurement of large numbers of particles to accurately characterize distributions.  More dispersed samples require particularly large numbers of measurements, because an exceedingly nonlinear scaling of detection sensitivity with particle size readily skews results to favour larger species \cite{hassellov2008nanoparticle}.  This presents a particular challenge because these conventional techniques for measuring nanoparticle size distributions are labour intensive, and thus have low throughput.  Progress in many areas of nanotechnology requires a faster, lower-cost and more accurate measure of particle size distributions in solution, particularly for diameters smaller than 20 nm.  

Interference backscattering microscopy (iSCAT) offers promise as a direct high-throughput probe of nanoparticle size distributions.  This technique measures the difference in refractive index between a particle and the medium. The light scattered from the particle interferes with the specular reflection formed at the interface with the microscope coverslip forming a background-free signal that scales linearly with the particle mass.  In 2006, Sandoghdar and coworkers obtained iSCAT images of single gold nanoparticles with diameters as small as 5 nm \cite{Jacobsen2006}.  The Kukura group established iSCAT as a tool to track the dynamics of macromolecules tagged with gold nanoparticles of similar size \cite{Ortega-Arroyo2012,OrtegaArroyo2016}.  iSCAT offers many strengths, including easy sample preparation for solutions with small volumes (10$\mu L$) and low concentrations (in the nM range), single-particle tracking, with mean size and distribution statistics independent of viscosity and temperature.  Here, we demonstrate a tailored instrument configuration, paired with a high-speed image-processing strategy, optimized for real-time particle tracking, that determines accurate size distributions in nominal 5 nm, 10 nm and 15 nm colloidal gold nanoparticle solutions, automatically sensing and classifying at a rate as high as 4,000 particles per minute.

\subsection*{Principles of interferometric scattering (iSCAT) mass photometry}

Several excellent reviews and protocols detail the background and operating principles of iSCAT microscopy \cite{Ortega-Arroyo2012,OrtegaArroyo2016,Cole2017,young2019interferometric,Taylor2019}.  {We provide a brief overview as context for the novel experiments described here, which sense the mass distributions of gold nanoparticles by direct count.  iSCAT operates by detecting the interference of light scattered by nano-object with that of a reference reflection. The iSCAT microscope detects  backscattered light with an intensity determined by:}

\begin{equation}
    I_{\text{det}} = 
    \left|E_i\right|^2[r^2 + s^2 + 2rs\cos{(\Delta\phi)}]
    \label{eqn:iscat_general}
\end{equation}

\noindent where $I_{\text{det}}$ is the intensity of light, $E_i$ is the incident electric field, $r$ is the fraction of $E_i$ backscattered by the reference field, $s$ is the fraction of $E_i$ scattered by a particle in solution, and $\Delta\phi$ is the phase difference between the reference and scattered light. In practice, the reference field is produced by the index of refraction difference between the exit of the coverslip ($n_{\text{glass}}=1.52$) and the sample medium ($n_{\text{water}}=1.33$). In a typical iSCAT configuration, the amount of light reflected by the coverslip dramatically overwhelms that which is weakly scattered by a particle in solution ($r^2\gg s^2$). Focusing the microscope on a nano-object bound to the coverslip surface maximizes destructive interference between the reference and scattered light ($\Delta\phi = \pi$). The object appears as a dark spot atop the constant reference field. This reduces Eq (\ref{eqn:iscat_general}) to $ I_{\text{det}} = \left|E_i\right|^2[r^2 - 2rs]$. 

Contributions to the measured intensity of the reference field ($I_{\text{ref}} = \left|E_i\right|^2r^2$) include, but are not limited to, backscattered light from the coverslip-sample medium interface, spurious reflections from optical elements in the detection path, fixed-pattern noise on the camera, fluorescence from the objective immersion oil, and non-uniform illumination. Because these contributions are both invariant to lateral sample translation and time independent, they can be isolated and removed by flat-fielding the image. This background normalization then defines the iSCAT contrast as follows:

\begin{equation}
    C_{\text{iSCAT}} = \frac{I_{\text{iSCAT}}}{I_{\text{BG}}} = 1-\frac{2s}{r}
    \label{eqn:contrast_iscat}
\end{equation}

\noindent where $I_{\text{BG}} = \left|E_i\right|^2r^2$. The scattering term $s$ is described by Mie scattering theory, where the diameter of the scattering object is comparable to the wavelength of incident light ($\lambda \approx D$) \cite{Ortega-Arroyo2012}. The magnitude of $s$ scales linearly with the volume of the particle\cite{Fan2014,Young2019}. The measured contrast of particles adhered to the coverslip surface is then described as follows:

\begin{equation}
    C_{\text{iSCAT}} = 1-{c(\varepsilon,r)D^3}
    \label{eqn:contrast_iscat2}
\end{equation}

\noindent where $c(\varepsilon,r)$ is a constant, incorporating the reflection constant $r$, the experimental detection efficiency, and the dielectric constants, $\varepsilon$, of the particle and surrounding medium\cite{Ortega-Arroyo2012}. For particles of known density, the scattering term scales linearly with the mass of the particle ($m = \rho D^3$). Substituting this equivalency into Eq (\ref{eqn:contrast_iscat2}) yields $C_{\text{iSCAT}} = 1-c(\varepsilon,r,\rho)m$.

Careful control of experimental conditions can increase sensitivity to a point limited by shot-noise-induced fluctuations in the laser light incident on the CMOS camera. For a shot-noise-limited experiment, the signal-to-noise ratio (SNR) depends on the square root of the number of photoelectrons accumulated by the detector ($\text{SNR} \propto \sqrt{N}$).  Raising the laser power increases the number of detected photoelectrons, and this can improve the SNR.  But, saturation limits this strategy whenever the electron counts grow to exceed the pixel well-depth of the CMOS detector anywhere in the image.  

Contributions from the reference and scattered fields spatially separate near the back focal plane of the iSCAT objective, where the placement of a partially reflective mask can attenuate the amplitude of the reference field to a magnitude comparable to that of the scattering object\cite{Cole2017,Supp}.  This attenuation is essential for the detection of small, weakly-scattering objects.  

\section*{Experimental Methods}

\subsection*{Materials and iSCAT sample preparation}

All measurements reported here have used citrate stabilized Au nanoparticles samples as received from Sigma Aldrich (Product numbers 741949 (5 nm, 90 nM), 741957 (10 nm, 10 nM), 777137(15 nm, 2.7 nM)). Stock solutions were diluted with Mili-Q deionized water to yield 1.0, 0.5 and 0.1 nM deposition solutions, respectively.  Measurement began with the introduction of a diluted gold nanoparticle sample by micro-pipet to a 400 $\mu$L sample chamber fitted to the stage of the inverted iSCAT microscope (see below).  

A coverslip forms the bottom of the sample chamber and the window through which laser light illuminates the sample and the iSCAT optical train collects the scattered light.  Present measurements used high-precision poly-\textsc{D}-lysine coated coverslips for super-resolution microscopy (Neuvitro Corporation, GG-18-1.5H-PDL).  We use manufacturer-sterilized and vacuum sealed coverslips as delivered, without further treatment.

\subsection*{Transmission Electron Microscopy}

We acquired TEM images of gold nanoparticles deposited by dip coating onto a 3mm formvar copper TEM grid (Ted Pella: 01700-F, 200 mesh) using the Hitachi H-7600 transmission electron microscope in the UBC Bioimaging Facility.  The images were recorded using an accelerating voltage of 80kV at 200k magnification with 800ms exposure time at a resolution of 0.40nm per pixel.  We size-analyzed a total of 35 images (18 for 15nm, 5 for 10nm, 12 for 5nm) using a custom program coded in MATLAB 2020A. The software implements a two-stage circular Hough Transform\cite{Yuen1990,DAVIES2005} algorithm on a binarized image to identify the location and radius of each individual particle in the field of view.  {(See below and Supplemental Materials\cite{Supp} for details.)}. 

\subsection*{iSCAT instrumentation}

Figure~\ref{fig:instrument} details the complete microscope configuration used for gold nanoparticle mass photometry experiments. Custom software written in LabVIEW 2018mcollects image frames from two CMOS cameras, one for iSCAT imaging the other for automatic focus control.   

\begin{figure}[ht!]
    \centering
    \includegraphics[height=0.5\textwidth]{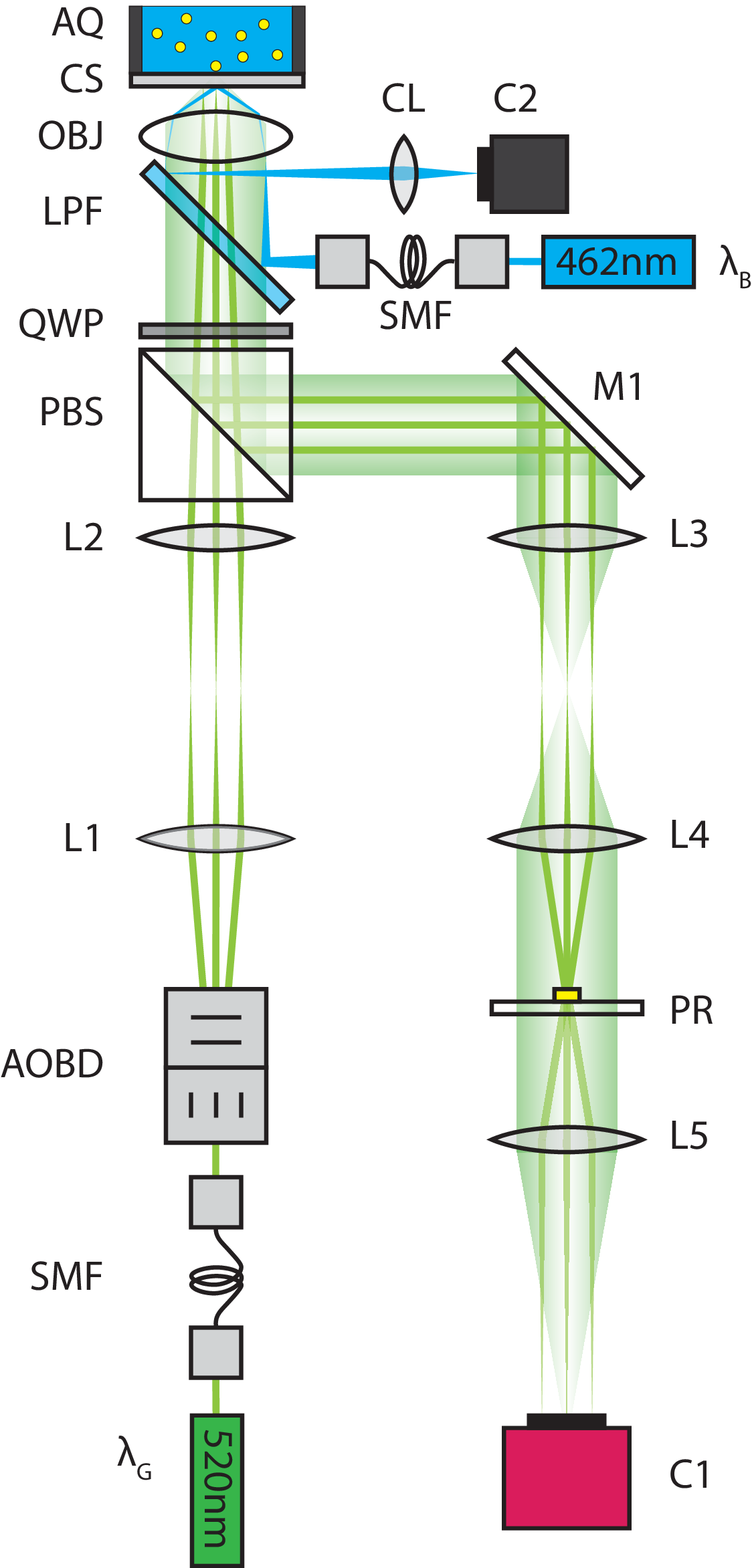}
    \caption{\footnotesize{Schematic diagram of the iSCAT microscope. The iSCAT channel operates in the green (520nm) and the auto-focus in blue (462nm).  Components include 520 nm and 462nm lasers ($\lambda_G$, $\lambda_B$) single mode fibers (SMF), $4f$ telecentric lens systems (L1, L2 and L3, L4),  acousto-optic beam deflector (AOBD), objective (OBJ), coverslip (CS), aqueous sample chamber (AQ), quarter wave-plate (QWP), polarized beam-splitter (PBS), partial reflector (PR), CMOS cameras (C1, C2), long pass filter (LPF), and cylindrical lens (CL).
    }}
    \label{fig:instrument}
\end{figure}

In the iSCAT channel, the output of a broadband 520 nm diode laser ($\lambda_G$: LaserTack LDM-520-1000-C) exits a single mode fibre (SMF: Thorlabs P1-488PM-FC-2) and passes through acousto-optic beam deflectors (AOBD: Gooch \& Housego R45100-5-6.5DEG-51-X/Y, MLV050-90-2AC-A1).  Here orthogonal TeO$_2$ crystals with LiNbO$_3$ slow shear wave transducers raster light collimated by a backwards facing objective (Olympus PlanN 20x / 0.40 NA) over a rectangular area. The beam propagates through a $4f$ telecentric lens system (L1,L2: f = 400mm) and quarter wave-plate (QWP: Thorlabs WPQ10M-532) to under-fill an infinity-corrected 1.42 NA oil immersion objective (OBJ: Olympus PlanApoN 60x / 1.42 NA / Oil / $\infty$ / 0.17). 

Back-scattered light from the sample and coverslip (CS: Neuvitro GG-18-1.5H-PDL) returns along the incident path.  Passing through the quarter wave-plate a second time rotates its polarization a total of 90 degrees, optimizing its deflection by a polarized beam-splitter (PBS). 

The backscattered beam then travels through a second $4f$ telecentric lens system (L3,L4: f = 400mm) which spatially separates the reference and scattered components.  A partial reflector mask (PR) positioned at the conjugate focal plane of the AOBDs selectively attenuates the reference light to match the intensity of the light scatted by sample particles. A CMOS camera (C1: Photonfocus MV-D1024E-3D01-160-CL-12, National Instruments PCIe-1433) with a $9.12 \times 9.12 ~\mu$m field of view detects each image. 

In the auto-focus channel, a 462nm diode laser ($\lambda_B$: LaserTack LDM-462-1400-C) exits a single mode fibre (SMF: Thorlabs P1-488PM-FC-2) A backwards facing objective (Olympus PlanN 4x / 0.10 NA) collimates this light and steers it to the objective by a 488nm dichroic long pass filter (LPF: Semrock Di03-R488-t3-25x36). Lateral translation of the LPF along the optical axis preserves the focus of this beam on back focal plane of the high-NA objective, displaced from its centre.  The resulting strong deflection of this beam causes a total internal reflection on the coverslip.  This reflected light is collected by the high-NA objective.  Any translation of the sample along the $z$-direction laterally translates the position of the reflected light focused by a cylindrical lens (CL: f = 400mm) on the Autofocus CMOS camera (C2: Point Grey GS3-U3-51S5C-C).  The microscope data system monitors the position of this signal to control the $z$ position of the iSCAT stage.\cite{Supp}

\subsection*{Sample stage}

\begin{figure}[ht!]
    \centering
    \includegraphics[height=0.3\textwidth]{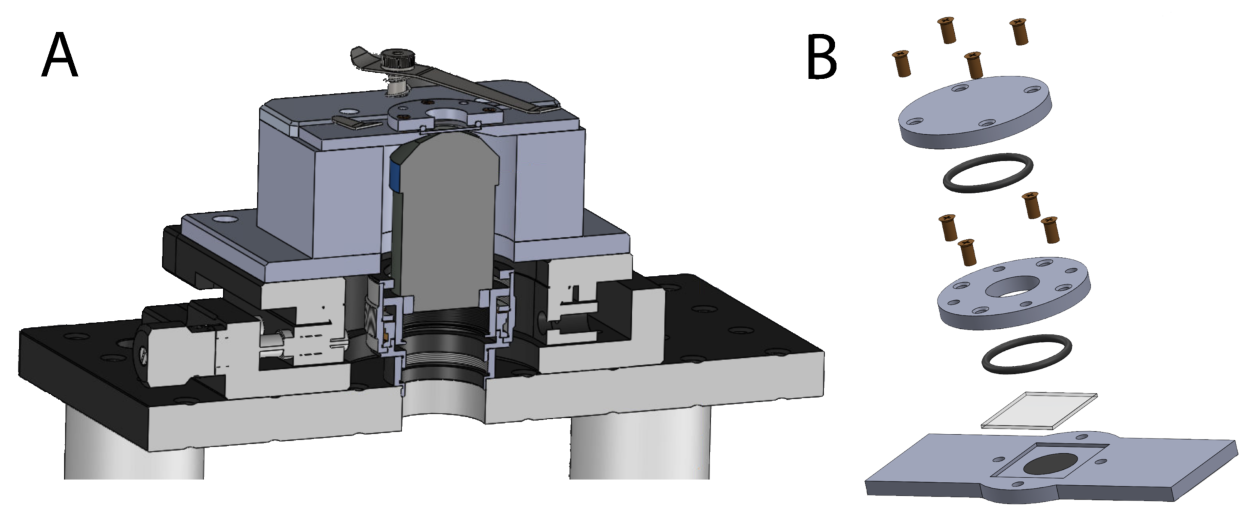}
    \caption{\footnotesize{(A) Cross sectional view of the microscope stage assembly. (B) Exploded representation of the aqueous sample holder. 
    }}
    \label{fig:stage}
\end{figure}

Figure \ref{fig:stage} shows a cross-sectional drawing of the iSCAT stage, together with an exploded view of the liquid sample chamber sealed at the bottom by a functionalized silica coverslip.  
A combination of mechanical and piezoelectric actuators control three-dimensional motion.  A mechanical stage (OWIS KT 90 XY) provides 20 mm of travel in the $x,y$ directions using manual actuators. A closed-loop piezoelectric stage (Piezosystems Jena Tritor 102 XYZ) provides 80 $\mu$m of travel in the $x,y$ and $z$ directions with a precision of $\pm 5$ nm.  The stage assembly accommodates a standard 18 mm coverslip sealed by a 16 mm i.d. viton o-ring in a stainless steel flange.

\subsection*{Image acqusition}

A purpose-written LabVIEW user interface operates the microscope, while custom codes written in MATLAB 2020A perform all data post processing.  With the camera operating at 500 fps with 2ms exposure time, a ratiometric stack of  $n_R = 50$ frames yields a temporal resolution of 10 Hz.  Choosing a region of interest (ROI) set to $304 \times 304$ pixels, each with a diameter of 30 nm defines a field of view on the detector of 9.12 x 9.12 $\mu$m.  The pixel diameter sets the limiting spatial resolution.  Normally, we use $2 \times 2$ pixel binning to generate a  $60 \times 60$ nm super-pixel, which increases the well depth by a factor of 4, from 200 to 800 kilo-electrons.  The AOBD oscillates the laser illumination to write a square on the entrance aperture of the objective at frequencies of 78 kHz and 80 kHz in the $x$ and $y$ directions, respectively.  We acquire and process all data on a computer with an Intel i7-4930K CPU at 3.40 GHz with 64 GB DDR3 RAM.

\subsection*{Ratiometric Processing}

A small gold nanoparticle presents a scattering amplitude no greater than the surface roughness typical of a coverslip. This necessitates accurate background noise identification and suppression. Unfortunately, the coverslip roughness pattern moves with lateral translation of the sample, and thus simple flat-fielding techniques cannot remove it \cite{OrtegaArroyo2016}.  In a case like this, time-dependent variations in the image in time can suppress a large static background to reveal a comparatively small transient signal.  This approach serves with particular effectiveness in particle binding experiments, such as that of gold nanoparticles in solution binding to the surface of a coverslip. 

Ratiometric image processing resolves particle binding events by balancing images that precede a particle landing event with those that follow the event \cite{Cole2017}.  This simple image processing technique isolates the landing events of individual nano-sized objects atop a shot noise-limited background. Particle landing events occur rapidly, on the order of a few milliseconds, so it is crucial to record video at the maximum possible frame rate. The present work optimizes the quality of the ratiometric signal by selective attenuation of the reference beam via the transmission of the partial reflector, and setting the laser illumination power to nearly saturate the CMOS camera at its maximum frame rate of 1kHz.

We ratiometrically isolate individual gold nanoparticle landing events by post-processing streams of raw CMOS images.  For each experiment, the application of a 210 nm low-pass Gaussian blur to every image in the full stack, $x(j)$ for $j = 1,2,...n_{\text{imgs}}$, acts as a low-pass filter, flattening all objects smaller than roughly twice the point-spread function.  This creates a parallel set of images, $x_{\text{LP}}(j)$ for $j = 1,2,...n_{\text{imgs}}$, representative of the background.  Dividing every raw image in a measurement by its low-pass filtered counterpart, $ {x(j)}/{x_{\text{LP}}(j)} = x_{\text{HP}}(j)$, yields a stack of high-pass filtered images, exposing all features smaller than 500 nm.  

Ratiometric processing seizes on the time-varying signal of an arriving particle to capture its background-free image.  To begin, we define a time-binning window, $N_R$, chosen to specify a number of high-pass filtered frames during which no two landing events overlap within the resolution of the point-spread function.  For our purposes, $N_R =50$, which at 500 Hz imaging defines a period of 0.1 s.    

To calculate a ratiometric image $\overline x_R(k)$, we examine the ratio of two sequential averaged windows comprising 50 frames each: $\overline x_R(k) = \overline x_2(k)/\overline x_1(k)$, where $\overline x_1(k)$ defines the average of frames 1-50, and $\overline x_2(k)$ figures the average of frames 51-100.  If no particle binding event occurs during the second sequence of frames, this ratio of images simply cancels to a field of values very close to 1.  The scattering signal produced by a particle binding event that occurs as late as frame 100 perturbs the ratiometric image, darkening or lightening it, depending on the relative phase of this signal.  In most cases, the phase of the light scattered from a landing particle interferes destructively with the reference signal, causing the particle to appear as a dark spot.

\begin{figure}[ht!]
    \centering
    \includegraphics[width=.65 \textwidth]{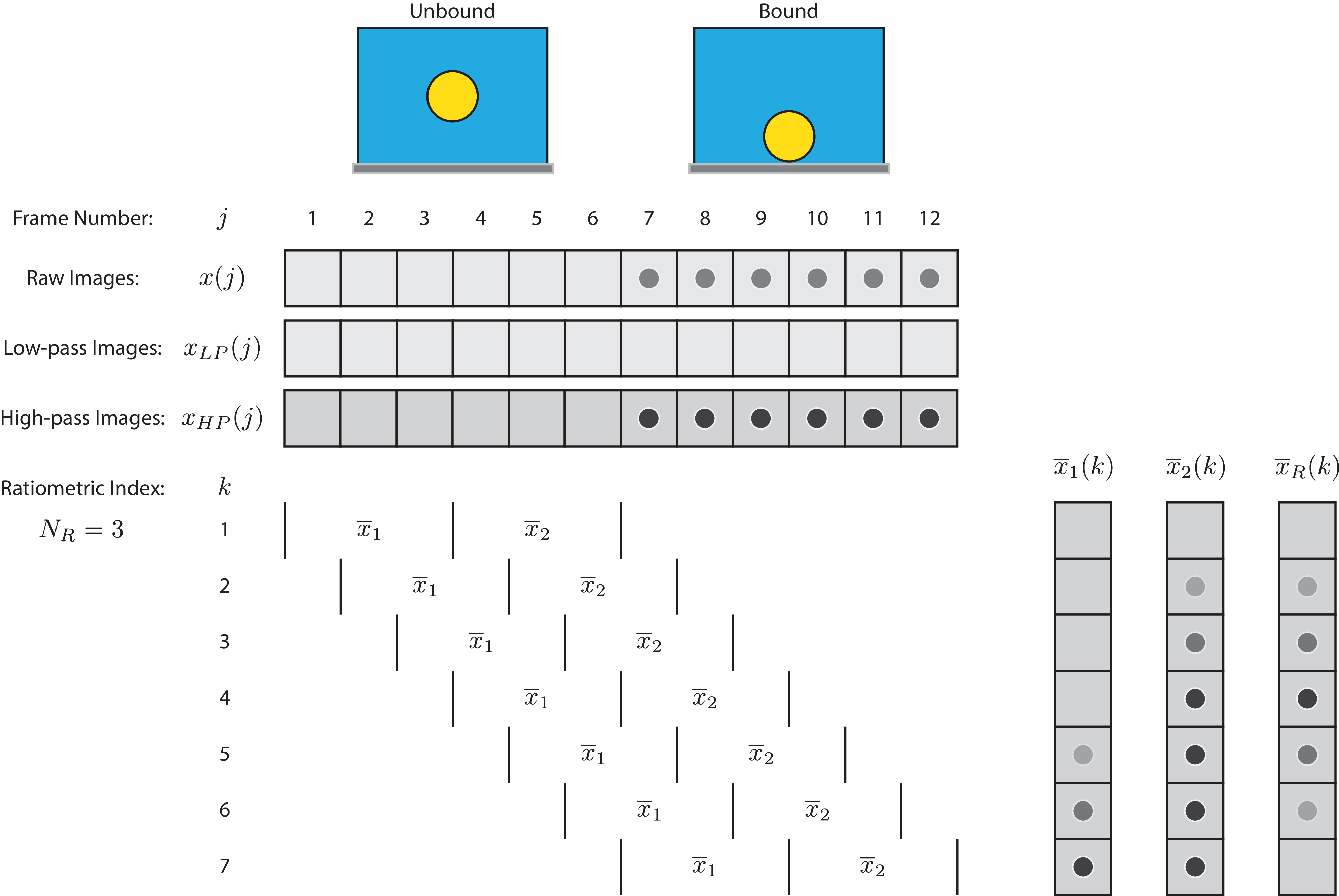}
    \caption{\footnotesize{Ratiometric image processing flowchart. A gold nanoparticle initially unbound in solution in frames 1-6, binds in frames 7-12.  An adjacent moving windows step through the images, defining $\overline x_1(k)$ and $\overline x_2(k)$, which combine to yield the ratiometric image $\overline x_R(k)$.}}
    \label{fig:ratiometric_diagram}
\end{figure}

Figure 3 schematically illustrates the ratiometric processing algorithm for a sequence of $j=12$ frames.  Here we see particle binding landing in frame $j=7$.  This example chooses $N_R =3$.  We progressively move the pair of averaged windows, $\overline x_1(k)$ and $\overline x_2(k)$ through a sequence of $k=7$ steps.  In step, $k=1$, both windows average frames that contain no new particles, and the ratio, $\overline x_R(1) = \overline x_2(1)/\overline x_1(1)$ defines a field with a near-uniform value very close to 1.   As $k$ advances through the sequence $(k=2...4)$, the particle first appears only in $\overline x_2(k)$,  and the ratiometric images, $\overline x_R(k)$, show a spot with growing contrast.  Starting with step $k=5$, the particle signal appears in both $\overline x_1(k)$ and $\overline x_2(k)$.  The contrast begins to cancel and the ratiometric image of the particle fades into the background.  

Note that this distinctive, one-sided variation of contrast with time characterizes the signal of a particle that binds permanently to the coverslip on the timescale of the measurement.  A particle that absorbs and then desorbs in a period of fewer than $N_R$ frames produces a decidedly different two-sided time-dependent contrast.  During the time when its signal is captured in frames that define $\overline x_2(k)$, a destructively interfering particle appears with a contrast value less than one.  Then, later in time, when the frames move forward to contain the particle signal wholly in $\overline x_1(k)$, the signal appears with a contrast value greater than one. 

{Differential (subtractive) processing suppresses the stationary background in a similar way to form a field distributed about a value of 0.  But, this subtraction of approximately equal numbers can lead to numerical underflow.  The division of approximately equal numbers in ratiometric processing helps circumvent this, because it consistently defines the background as a field of positive numbers near 1.  Most high-resolution images of small nanoparticles and light macromolecules use a ratiometric approach.\cite{OrtegaArroyo2016,Cole2017,Taylor2019,Supp} }

\subsection*{Single Particle Tracking}

{We refer to the automatic identification of dark circles as they appear in a sequence of ratiometric frames as particle tracking.}  A circular Hough transform (CHT) recognizes and locates particles that conform with a set of defining criteria.  For the present study, these include a diameter between 180 and 480 nm, and a contrast darker than the background value of 1.  For every particle in every frame, the tracking routine logs the location in the image, $(x,y)$, the diameter, a sub-image containing all of the pixels of the particle found by CHT, a two-dimensional Gaussian fit of the particle sub-image, and the \textit{x} and \textit{y} image cross-sections centred on the particle. The tracking routine judges circles found in subsequent frames as the same particle if the \textit{xy} position varies by less than 2 pixels. 

Filters applied to this library of particles remove false events.  We reject particle images that conform with eccentric two dimensional Gaussians, $0.75 < \sigma_x/\sigma_y < 1.25$.  Bound particles must appear in a sequential number of ratiometric frames that exceeds $N_R$.  We reject any particle with a centre position within 720 nm of any edge of the image.  This reduces the effective field of view from $9.12 \times 9.12$ $\mu$m to $7.68 \times 7.68$ $\mu$m.  Finally, we fit two linear functions of equal but opposite slope to the leading and trailing sides of the Gaussian amplitude as a function of frame number for each particle.  We require the goodness of fit for these two lines to conform with $r^2_{x,y} \ge 0.98$. We log events that meet all of the above criteria in a particle inventory for size and kinetic analyses.  We apply a very similar but static image-processing strategy to gauge nanoparticle diameters in the TEM micrographs.

\section*{Results}

This work examines the particle sizes found in commercial samples of colloidal gold nanoparticles with nominal diameters of 5, 10 and 15 nm.  We compare size distributions derived from the analysis of many single-particle images obtained by transmission electron microscopy (TEM) with iSCAT measures acquired dynamically by high-throughput particle counting.  

\subsection*{Transmission electron microscopy of gold nanoparticle standards}

\begin{figure}[ht!]
    \centering
    \includegraphics[width=.77 \textwidth]{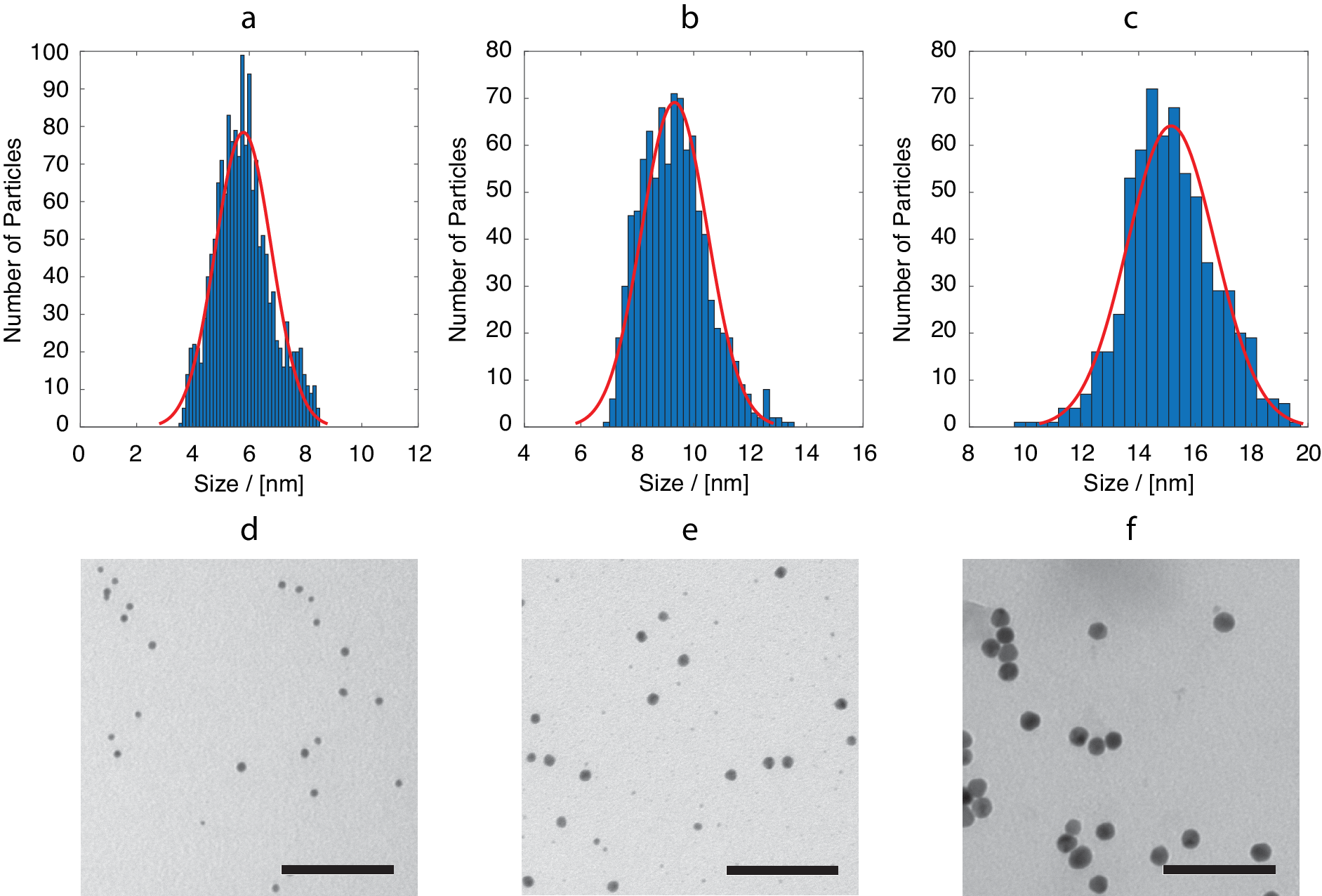}
    \caption{Representative TEM images recorded for nominal [a] 5nm , [b] 10nm, and [c] 15nm gold nanoparticle solutions. Scale bars are 100 nm.}
    \label{fig:TEM}
\end{figure}

Fig.~\ref{fig:TEM} shows representative portions of three TEM images recorded for samples prepared from colloidal gold nanoparticles with nominal diameters of 5, 10 and 15 nm.  
{{CHT machine-vision image-processing routines applied to these and companion micrographs in full frame yield measured diameters for a number of particles at each nominal diameter that ranges from 642 (15 nm) to 1,661 (5 nm).  Note that the discrete count as a function of exact size conforms well with a Gaussian distribution for a nominal particle size of 15 nm.   However, a particle boundary threshold that both distinguishes smaller particles from the noise and captures the Gaussian distribution of larger particle sizes detects fewer particles than expected in the lower-diameter tails of the 5 and 10 nm distributions.  Despite this evident limitation in dynamic range, the CHT processing of TEM micrographs produces direct counts with well-matched means and median diameters:  $5.73-5.68$ nm,  $9.32-9.24$ nm, and $15.15-15.05$ nm from the analysis of 1661, 929, and 642 particles respectively.\cite{Supp}  Respective Gaussian widths of 1.0, 1.2 and 1.6 nm conform with the manufacturer's declared size precision of $\sim10$\%.}}

\subsection*{Ratiometric imaging of single gold nanoparticles}

\begin{figure}[ht!]
    \centering
    \includegraphics[width=.7 \textwidth]{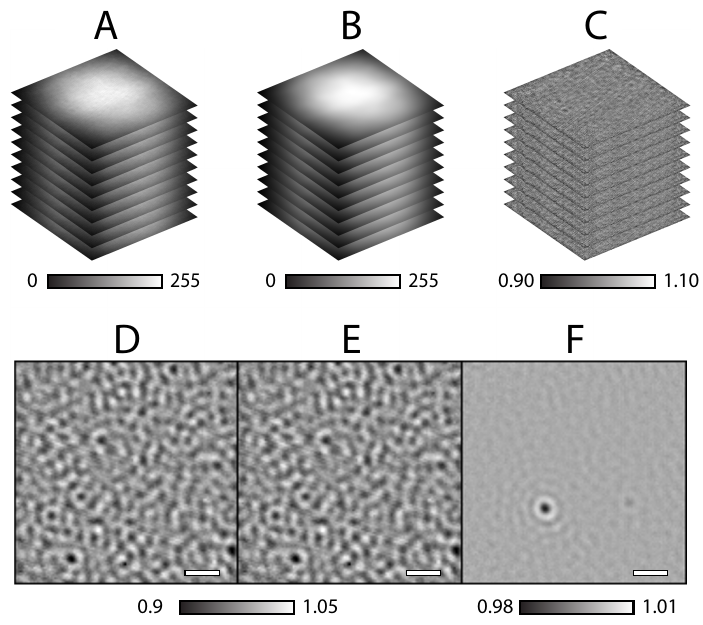}
    \caption{\footnotesize{(A) A schematic stack of raw  0.1 megapixel images representing the sequence of frames recorded in an iSCAT experiment.  (B) The same sequence of frames after 210 nm Gaussian low-pass filtering.  (C) The matching sequence of high-pass filtered images formed by dividing each raw frame in (A) by its low-pass filtered counterpart in (B).  (D)  Average of 50 frames immediately before the landing of a 10 nm gold nanoparticle.  (E)  A sequence of 50 frames that begins when a nanoparticle lands from solution and binds to the coverslip.  (F)  Ratiometric image of this nanoparticle at its point of maximum contrast, formed by dividing image (E) by image (D).  } }
    \label{fig:Ratiometric}
\end{figure}

The acquisition of iSCAT images resolving single nanoparticle landings requires a series of steps in a strategy of high-throughput image processing.  Figure \ref{fig:Ratiometric} illustrates this strategy by a sequence of CMOS camera images beginning with the raw output and ending with the ratiometric image of a single 10 nm gold nanoparticle.  

Figure \ref{fig:Ratiometric}A shows a representative stack of raw, 0.1 megapixel images as acquired at 500 Hz and stored over the course of an experiment.  The post processing of the data from an experiment begins with the application of a 210 nm Gaussian blur as described above.  This generates a parallel sequence of low-pass filtered of images, represented by Figure \ref{fig:Ratiometric}B.  Dividing each raw image A by its low-pass filtered self.  This yields a stack of high-pass filtered images, Figure \ref{fig:Ratiometric}C, in which spatially sharp structure sits atop a pseudo flat-field in which the low-frequency structure has been smoothed to an amplitude of 1.  

This stack of flat-fielded images serves as input for ratiometric processing to uncover the signals of single particles as they land.  Figure \ref{fig:Ratiometric}D shows the average of flat-fielded 50 frames combined to represent an unbound state, $\bar x_1$, before the landing of a new nanoparticle.  The centre frame, Figure \ref{fig:Ratiometric}E, shows a stack of 50 sequential images obtained immediately after the landing of a new nanoparticle.  This nanoparticle exists in every frame of the centre image average, $\bar x_2$,  but in none of the images averaged to form the left frame.  Dividing the left frame into the centre frame produces the image shown on the right, in which the nanoparticle is very clearly visible.  Notice the contrast scale becomes more sensitive in the ratiometric image (0.98 - 1.01) compared to the averaged frames alone (0.9-1.05).

\subsection*{Batch tracking and classification of nanoparticle landings}

\begin{figure}[ht!]
    \centering
    \includegraphics[width=.65 \textwidth]{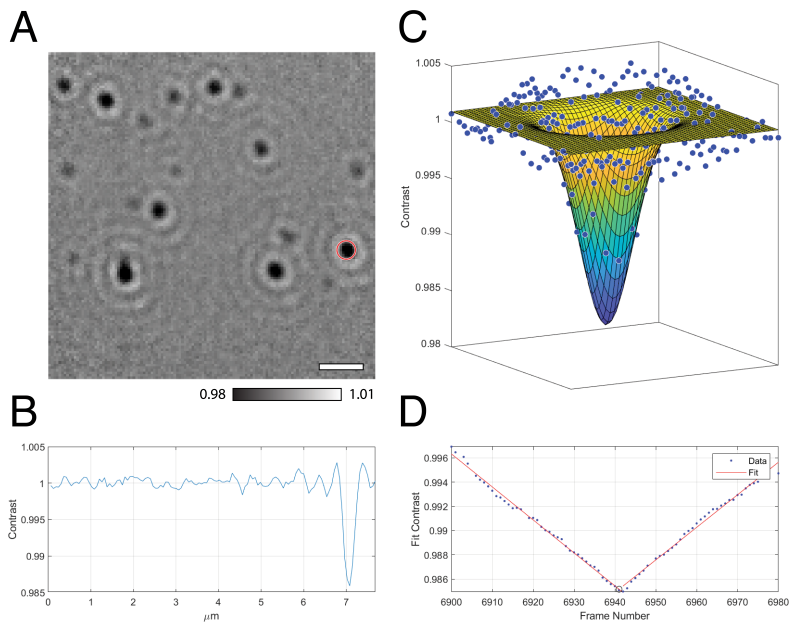}
    \caption{\footnotesize{Particle tracking characteristics for a single 10 nm gold nanoparticle. (A) Ratiometric image at frame 6942 with particle of interest circled in red. Scale bar is 1 $\mu$m. (B) $x$-profile cross section of the image in (A) with SNR of 7.35.  (C) two-dimensional Gaussian fit to the point spread function of the gold nanoparticle at the precise frame it binds (6942). (D)  two-dimensional Gaussian amplitude fitting parameter as a function of frame number. As the gold nanoparticle approaches the cover-slip, the image becomes darker until the precise moment it binds to the glass surface. It then becomes lighter until it vanishes into the background. The intersection of two linear fits determines the precise time which the binding event occurred, and its associated contrast value.}}
    \label{fig:aunp_10nm_single_particle}
\end{figure}

We have optimized the current configuration of this instrument and its operating software to capture and classify every nanoparticle that lands from solution onto the iSCAT microscope coverslip.  Figure \ref{fig:aunp_10nm_single_particle} shows image number 6,942 in a sequence obtained by the ratiometric processing of 20,000 consecutive frames recorded at 500 fps over a period of 40 seconds for a 0.47 nM solution of 10 nm gold nanoparticles.  

During this time, particle signals appear as a depletion signal that grows linearly from the $A=1$ background over an interval of about 50 frames, and then fades at the same rate.  At the particular point of Figure \ref{fig:aunp_10nm_single_particle}A, the Hough transform algorithm has identified 18 particles.  The one particle marked by a red circle has reached the maximum ratiometric contrast displayed in Figure \ref{fig:aunp_10nm_single_particle}D.  A two-dimensional Gaussian fit to the local distribution of extinction values in the space of CMOS super pixels determines the Gaussian amplitude, $A = 0.9825 \pm 0,0013$ and position $x,y = 7,026 \pm 35$, $4,690 \pm 16$ nm and width $\sigma_x, ~\sigma_y = 115.64 \pm 11.8$ nm, $140.22 \pm 10.69$ nm with 95\% confidence bounds.  Figure \ref{fig:aunp_10nm_single_particle}B plots a section of the ratiometric image cut through this particle at a fixed $y=4,690$ nm.  For this frame, the measurement time is 14 seconds.  At this point, particles are landing at a rate of 51 s$^{-1}$. 

Point-spread functions with Gaussian widths of $\sigma = 316 \pm 6$ nm adequately describe all the ratiometric signals produced by particles with diameters of 5, 10 and 15 nm.   Figure \ref{fig:aunp_10nm_single_particle}C shows a three dimensional representation of the Gaussian point-spread function fit to the signal for the particle highlighted in Figure 3A.   

\subsection*{Distribution of iSCAT contrast for nominal 5, 10 and 15 nm gold nanoparticles}

\begin{figure}[ht!]
    \centering
    \includegraphics[width=.6 \textwidth]{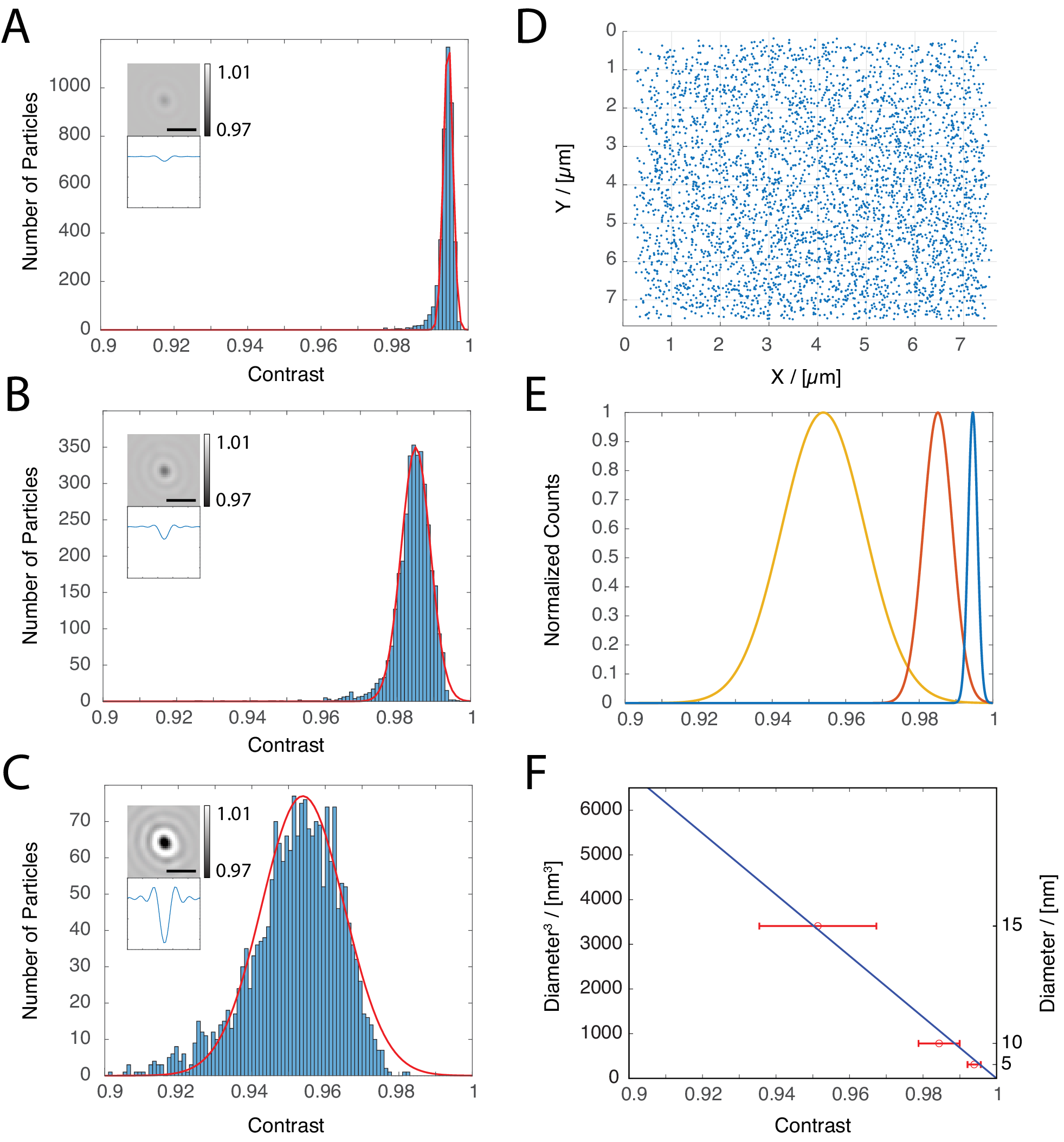}
    \caption{\footnotesize{(A-C) Histograms binning contrast values from single particle tracking coverslip binding experiments for (A) 5nm, (B) 10nm, (C) 15nm standards. Insets show representative iSCAT images (top) with cross sectional profiles (bottom). The contrast ranges from 0.97 to 1.01 with 1 $\mu$m scale bars. (D) Landing locations on the coverslip for the 5nm experiment. (E) Overlaid distributions fit to histograms A-C. (F) Calibration plot relating the average volume of the particle ($D^3$) determined by TEM to the contrast value measured by iSCAT. Error bars correspond the FWHM of each gaussian fit from A-C.}}
    \label{fig:rep_particles}
\end{figure}

We observe that the ratiometric contrast varies with the nominal particle size.  Figure \ref{fig:rep_particles} summarizes results for 5, 10 and 15 nm gold nanoparticles.  Histograms plotted as Figure \ref{fig:rep_particles}A, B and C display the number of landings versus the Gaussian amplitude fit to each particle point-spread function.  Insets in each of these figures picture the observed point-spread functions for each particle, obtained as an average of the peak signal observed for all particles of each size.  The image contrast increases with particle size.  However, the two-dimensional Gaussian fit yields the same full-width at half-maximum, $316 \pm 6$ nm for 5, 10 and 15 nm. Notice the Airy diffraction rings around each particle, which grow in prominence with increasing particle size. 

Figure \ref{fig:rep_particles}D plots the $x,y$ coordinates of each landing detected in the 80 second measurement that produced the 5 nm gold nanoparticle distribution pictured in frame A.  During this time, 4,196 detected particles landed within the $7.68  \times 7.68$ $\mu$m field of view (FOV).  By end of this experiment, gold nanoparticles cover about 0.2\% of the surface area of the coverslip.  Fewer particles land in the 10 nm experiment.  Here, we detect and classify 2,783 particles, covering 0.5\% of the FOV.  The 15 nm experiment combines the results of three experiments, classifying 781, 689 and 612 particles respectively, in each case occupying about 0.2\% of the FOV.  We fit these size distributions to the Gaussians superimposed on each histogram, and collected in Figure \ref{fig:rep_particles}E.

The fit of a Gaussian function to the ratiometric signal distribution for each gold nanoparticle determines peak contrast values of $C_{\text{5nm}} = 0.9946 \pm 0.1\%$, $C_{\text{10nm}} = 0.9851 \pm 0.1\%$, and $C_{\text{15nm}} = 0.9539 \pm 0.2\%$ with a 95\% confidence interval.  Figure \ref{fig:rep_particles}F plots the mean diameters determined by TEM for colloidal solutions of nominal 5, 10 and 15 nm gold nanoparticles versus the mean contrast measured for each of these solutions by iSCAT.  This correlation supports a regression model for $c(\varepsilon,r)$ in Eq (\ref{eqn:contrast_iscat}) relating measured iSCAT contrast to the diameter of any individual gold nanoparticle.   The resulting calibration function Eq.~\ref{eqn:contrast_calibration} provides a predictive model by which to determine size of an unknown gold nanoparticle from its measured iSCAT contrast with an RMS error of 2.73 nm. Notice the width the distribution increases with the size of the nanoparticle.

\begin{equation}
   { C_{\text{AuNP}} = 1-(1.460*10^{-5}\text{nm}^{-3})D^3}
    \label{eqn:contrast_calibration}
\end{equation}

\subsection*{Time evolution of gold nanoparticle deposition}

\begin{figure}[ht!]
    \centering
   \includegraphics[width=0.65\textwidth]{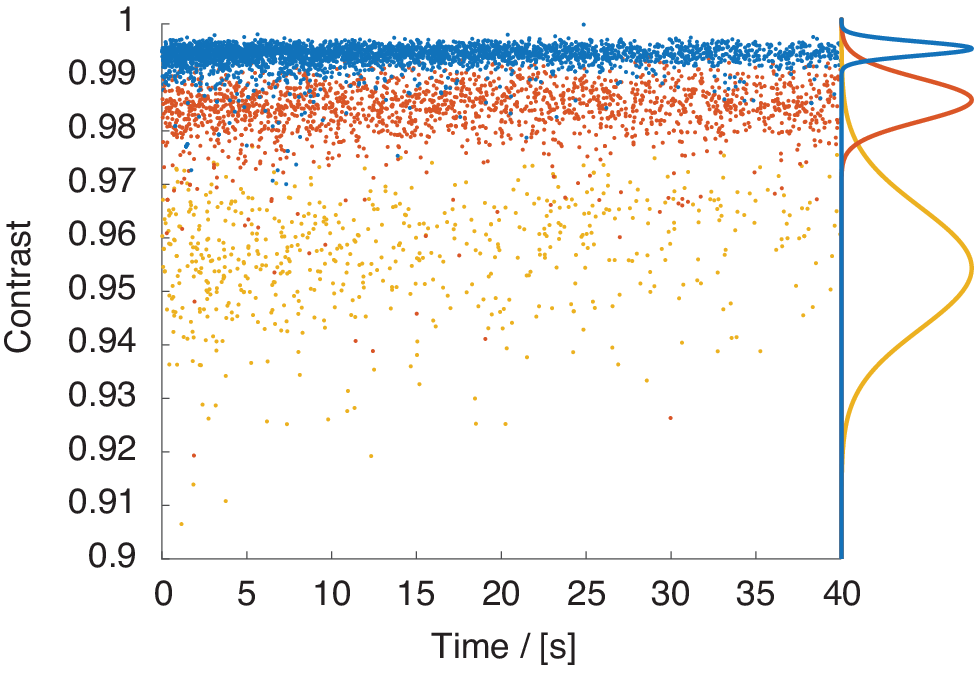}
    \caption{\footnotesize{Landing events detailed by contrast as a function of time for 5nm [blue], 10nm [red], and 15nm [yellow] gold nanoparticle standards.  During this period of 40 seconds, these solutions, with concentrations as indicated in Table \ref{tab:deposition}, deposit 3456, 1880 and 612 particles, respectively.  The cumulative counts fit to Gaussian distributions, as displayed on the right.  These curves skew slightly to larger particle size compared with the distributions accumulated over 80 s, as pictured in  Fig.~\ref{fig:rep_particles}E.}}
    \label{fig:kinetics}
\end{figure}

Figure \ref{fig:kinetics} plots the contrast observed as a function of time for landing events as they occur in the initial 40 seconds of experiments that record the deposition of 5, 10 and 15 nm gold nanoparticles on poly-D-lysine coated coverslips.  For all particle diameters, note how landing events occur with substantially greater frequency in the first 10 seconds of observation, and diminish substantially by the end of this observation period.  It also appears that larger particles land with greater frequency at earlier times for particle of all three diameters.

  \begin{table}[ht]
	\centering
	\begin{tabular}{rrrrr}  
 \caption{\footnotesize{{{Fractional coverage of gold nanoparticles deposited on poly-D-lysine coated cover slips in representative runs of 80 seconds  (5 and 10 nm), and 40 seconds  (15 nm).  58.98 $\mu$m$^2$ field of view (FOV) defined by the $7.68 \times 7.68$ $\mu$m central area of $304 \times 304$ CMOS pixels.  Total active area of $1.98\times 10^8$ $\mu$m$^2$ determined by the $\frac{5}{8}$ inch o-ring i.d.  }}}}
\label{tab:deposition}
  \multirow{2}{*}{\parbox{2 cm}{\centering Diameter\\ (nm)}}
&  \multirow{2}{*}{\parbox{3cm}{ \centering Area per Particle (nm$^2$)}}
&  \multirow{2}{*}{\parbox{3cm}{\centering Number in\\ FOV }}
&  \multirow{2}{*}{\parbox{3 cm}{\centering Covered Area \\  (nm$^2$)}}
&  \multirow{2}{*}{\parbox{2.5 cm}{\centering Fractional \\ Coverage}} \\ \\ \\
\toprule
5 \hspace{15 pt}    & 19.63  \hspace{25 pt}   &   4,196 \hspace{28 pt}   &   82388  \hspace{25 pt} &   0.002  \hspace{20 pt}  \\ 
10 \hspace{19 pt}  & 78.54  \hspace{25 pt}  &   2,783 \hspace{28 pt}   &   218576  \hspace{25 pt} &   0.005  \hspace{20 pt}  \\ 
15 \hspace{19 pt}  & 176.71  \hspace{25 pt}  &   571 \hspace{28 pt}   &   100904  \hspace{25 pt} &   0.002  \hspace{20 pt}  \\ 
	\bottomrule
	\end{tabular}
\end{table}

Table \ref{tab:time_depen} details the rate properties of gold nanoparticle deposition, as measured by a direct count of single landing events.  Here, we show results for single solutions of 5 and 10 nm particles for intervals of 80 seconds, and three measurements of deposition from colloidal solutions of 15 nm gold nanoparticles monitored for 40 seconds.  Note that in all cases studied, irreversible binding on the poly-D-lysine coated coverslips consumes only a small fraction of the particles in solution.  Yet, the rate of deposition, as monitored by direct count, diminishes over the time of observation by as much as an order of magnitude or more.  

  \begin{table}[h!]
	\centering
	\begin{tabular}{rrcrcrrr}  
 \caption{\footnotesize{Rate properties of the deposition of 5, 10 and 15 nm diameter gold nanoparticle on poly-D-lysine coated cover slips from 400 $\mu$L volume colloidal solutions with the concentrations indicated.  Note that after 40 seconds, bound nanoparticles occupy less than 0.5 percent of the coverslip area, depleting the concentration of nanoparticles in solution by no more than 7.8 percent, yet the measured rate of particle binding falls by as much as an order of magnitude or more.  
}}
\label{tab:time_depen}
  \multirow{2}{*}{\parbox{1.3 cm}{\centering Diameter\\ (nm)}}
&  \multirow{2}{*}{\parbox{2 cm}{ \centering   [AuNP]$_0$ \\(nM)\\}}
&  \multirow{2}{*}{\parbox{1.7 cm}{\centering  Total Number \\ $\times 10^{-11}$}}
&  \multirow{2}{*}{\parbox{1.5 cm}{\centering Elapsed \\Time (s)\\}}
&  \multirow{3}{*}{\parbox{1.7 cm}{\centering Number \\Landed\\ $\times 10^{-10}$}}
& \multirow{2}{*}{\parbox{1.4 cm}{\centering Percent Landed}} 
& \multirow{2}{*}{\parbox{1cm}{\centering Initial Rate \\ (s$^{-1}$)}} 
& \multirow{2}{*}{\parbox{1cm}{\centering Final Rate \\  (s$^{-1}$)}}   \\ \\ \\
\toprule
5 \hspace{11 pt}     &1.13 \hspace{20 pt}      &   2.75 \hspace{2 pt} &80 \hspace{12 pt}  & \hspace{1 pt}   1.41    &  5.1   \hspace{9 pt}    &  190  \hspace{3 pt}    &   10   \hspace{6 pt}    \\
10 \hspace{12 pt}  &0.47 \hspace{20 pt}      &   1.20 \hspace{2 pt} &80 \hspace{12 pt}  &   0.93   &   7.8   \hspace{9 pt}    &    70  \hspace{3 pt}    &  15   \hspace{6 pt}    \\
15 \hspace{12 pt}  &0.13 \hspace{20 pt}      &   0.33 \hspace{2 pt} &40 \hspace{12 pt}  &   0.19    &   5.8   \hspace{9 pt}    &   41  \hspace{3 pt}    &   4   \hspace{6 pt}    \\
	\bottomrule
	\end{tabular}
\end{table}

\section*{Discussion}

\subsection*{Comparison of TEM and iSCAT Distributions}

\begin{figure}[ht!]
    \centering
    \includegraphics[width=0.7\textwidth]{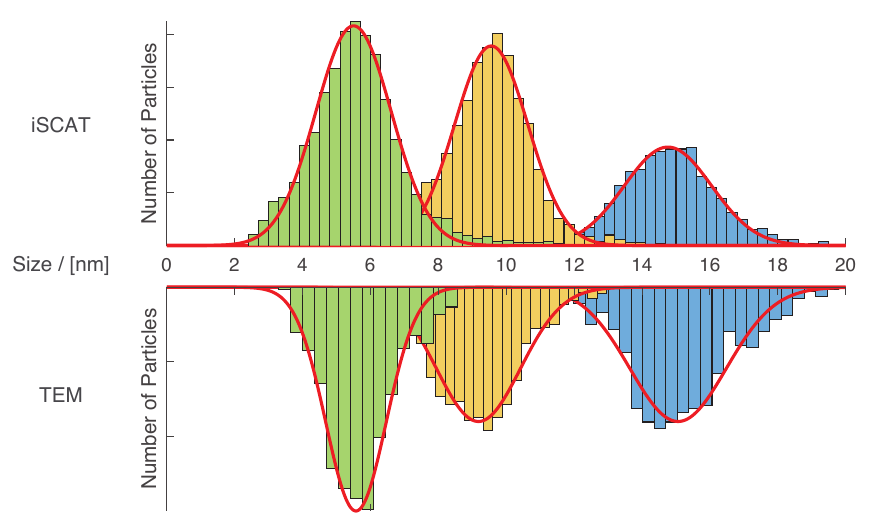}
    \caption{\footnotesize{Comparison of the size distributions determined from combined iSCAT experiments [top] and from TEM images [bottom]. Distributions for 5nm, 10nm, and 15nm samples are shown in green, yellow, and blue, respectively.}}
    \label{fig:hist_iscat_tem}
\end{figure}

Results presented here linearly relate the mean volume of gold nanoparticles measured by the analysis of TEM images to the mean contrast measured for the same nanoparticle solutions by the ratiometric imaging and analysis of individual iSCAT landings.  This relationship, plotted above in Figure  \ref{fig:rep_particles}F, determines an instrumental value for $c(\varepsilon,r)$, which enables the conversion of contrast values measured in the present iSCAT instrument configuration to nanoparticle diameters.  Figure \ref{fig:hist_iscat_tem} compares the distribution of nanoparticle diameters determined by the iSCAT identification and classification of particles drawn from solutions of 5, 10 and 15 nm gold nanoparticles to the distributions measured by processing TEM images of nanoparticles drawn from the same solutions.  

\begin{table}
\begin{tabular}{ cccccc } 
\hline
 & \makecell{Standard Size \\ (nm)} & \makecell{Number of \\ Particles} &  \makecell{Mean Size\\ (nm)}  & \makecell{Median Size \\ (nm)} & \makecell{Gaussian $\sigma$ \\ (nm)} 
  \\
\toprule
		&	5 	& 4,196 	& 5.60	& 5.53	& 1.4 		\\ 
iSCAT	&	10 	& 3,524 	& 9.53	& 9.55 	& 1.3 		\\
		&	15 	& 2,083 	& 14.89	& 14.82 	& 1.4 		\\  
\midrule
 		&	5 	& {1,661} 	& 5.73	& 5.68 	& 1.0 		\\ 
TEM		&	10 	& 929 	& 9.32	& 9.24 	& 1.2 		\\
		&	15 	& 642 	& 15.15	& 15.05 	& 1.6 		\\  
\bottomrule
\end{tabular}
\caption{\label{tab:tem_iscat_table} \footnotesize{Properties of the distribution of diameters determined by the size of nanoparticles measured directly by the processing of TEM images and from the calibrated contrast found for individual particle landings in iSCAT for solutions of 5, 10 and 15 nm gold nanoparticles.}}
\end{table}

{Figure \ref{fig:hist_iscat_tem} compares the particle size distributions determined by both iSCAT and TEM.  In both cases, the distributions reflect  cumulative measurements of exact diameters of particles sampled from nominal 5, 10 and 15 nm solutions.   iSCAT directly gauges the interferometric scattering of each gold nanoparticle.  The histograms plotted convert from the ratiometric contrast value, which scales with $D^3$, to diameter using the calibration function defined in Eq (\ref{eqn:contrast_calibration}).  Table \ref{tab:tem_iscat_table} summarizes the parameters of these distributions.  Note that the means and median values of these distributions match in keeping with normal distributions of particle size in the stock solutions.  The agreement between the widths of iSCAT and TEM diameter distributions adds weight to the linear fit of interferometric contrast to $D^3$ in Figure \ref{fig:rep_particles}F, and supports the concept of iSCAT as a means of nanoparticle mass photometry.\cite{Supp}  }

\subsection*{Dynamics of surface deposition from nanoparticle solutions}

Over a period of 80 s, gold nanoparticles deposit on poly-\textsc{D}-lysine coated coverslips from colloidal solutions of concentration from 1.0 to 0.1 nM to occupy substantially less than 1\% of the available surface area, and, under the present conditions, consume substantially less than 10\% of the initial nanoparticle concentration.  Yet, inspecting the time evolution of the deposition, as described by Figure \ref{fig:kinetics}, we can see that after only 40 s, the rate of deposition slows noticeably.  It also seems clear that deposition samples a distribution of contrast {that places a greater weight on larger particles at earlier times.  

{This observed fall off in deposition very clearly fails to conform with simple Langmuir kinetics.  The mass selectivity apparent in Figure \ref{fig:kinetics} holds a key, suggesting a limitation owing to the transport properties of nanoparticles in quiescent solutions.  For example, in the field of view, we count 5 nm gold nanoparticles landing with an initial a rate of 190 s$^{-1}$.  Extrapolating to the full $2 \times 10^8$ $\mu$m$^2$ contact area of the coverslip, the solution loses particles at an initial rate of $6.4 \times 10^8$ s$^{-1}$.  A Stokes-Einstein estimate of the diffusion coefficient of 5 nm nanoparticles predicts that such a rate of deposition will drive a non-equilibrium response in the solution.  Consider a first disk  of solution, 20 $\mu$m high, in immediate contact with the coverslip.  This disk initially contains about $2.7 \times 10^9$ particles Over the first one second, binding to the coverslip diminishes this number by about 27 percent.  This produces a concentration gradient with respect to the disk above, which drives a diffusion of particles into the depleted layer at a rate of $8 \times 10^7$ s$^{-1}$ -- nearly an order of magnitude slower than the rate of particle loss to binding on the coverslip.  We can express this differential transport in terms of the fractional degree to which diffusion from the upper level replenishes nanoparticles lost in the first second to binding on the coverslip.  For 5 nm particles, this approximate picture predicts a factional replenishment of 0.125.  The same model yields fractional replenishments of 0.065 and 0.036 for particles with the diffusion coefficients of 10 nm and 15 nm gold nanoparticles.  See the Supplemental Materials for details.\cite{Supp}  This effect of growing depletion with particle size explains the size selectivity that grows in time clearly evidenced in Figure \ref{fig:kinetics}.  

A fuller description of these non-equilibrium deposition kinetics will, of course, require a finer-grained consideration of volume elements and their variation in time.  But, this simple finite reckoning of deposition versus diffusion over a step of one second and distance of 20 $\mu$m shows how readily the irreversible binding of nanoparticles to an attractive surface can outpace diffusion, even for particles of the smallest size and highest diffusion coefficient.   }

\section*{Conclusions}

Results presented here demonstrate that interferometric scattering microscopy (iSCAT) can serve as a sensitive, high-throughput imaging methodology for mapping distributions of size in colloidal solutions of nanoparticles.  This protocol demands little sample preparation and requires only microlitres of solutions with concentrations no greater than the nano-molar range.  The high throughput of the image-processing and particle-tracking techniques introduced here offer sensitivity across a wide size distribution of nanoparticles with a dynamic range that faithfully samples rare instances of species in its leading and trailing edges.  Resolving single particle-coverslip binding events as a function of time, this work appears to have uncovered a transport-limited regime of gold nanoparticle binding kinetics.  

\begin{acknowledgement}

We thank the Natural Sciences and Engineering Research Council of Canada (NSERC) and Canfor Pulp Innovation for joining in a Collaborative Research and Development grant which provided support for this work. We also gratefully acknowledge equipment support from the Canada Foundation for Innovation and the British Columbia Knowledge Development Fund.

\end{acknowledgement}


\newpage

\renewcommand{\thetable}{S\arabic{table}}   
\renewcommand{\thefigure}{S\arabic{figure}}

\setcounter{figure}{0}
\setcounter{table}{0}

\begin{center}

{\Large \bf{Supplemental Materials\\
  Size distributions of gold nanoparticles in solution measured by single-particle mass photometry}}

{Luke Melo, Angus Hui, Matt Kowal, Eric Boateng, Zahra Poursorkh, Ed\`ene Rocheron, Jake Wong, Ashton Christy and Edward Grant}

{\it{Department of Chemistry, The University of British Columbia \linebreak Vancouver BC, Canada, V6T 1Z1 }}

 
\end{center}

\section*{Characteristics of the iSCAT optical train and data processing system optimized for measuring gold nanoparticle size distributions}

The following sections detail aspects of the iSCAT microscope optical train and data processing routines that proved most instrumental in automating the high-throughput measurement of gold nanoparticle size distributions.  

\subsection*{Attenuation of the specularly reflected reference beam by means of a partial reflector mask}

The second element of the 4f telecentric lens system (L3,L4: f = 400mm) focuses the reference beam specularly reflected by the cover slip onto the the conjugate focal plane of the acousto-optic beam deflectors (AOBDs).  There, a partially reflecting gold film, deposited to form a central 2 mm spot, attenuates the reference light to balance the intensity of the light scattered by sample particles.  

We have obtained partial reflector masks with a range of optical densities by means of special-purpose coating runs in the Nanofabrication Facility of the Stewart Blusson Quantum Matter Institute at UBC.  Using a custom stencil to define a 2 mm diameter spot in the centre of a 1mm thick optical flat (Thorlabs: WG11010), we deposit a 5 nm titanium adhesion layer followed by evaporated gold to a predetermined thickness, ranging from 25nm to 100nm.  Measurements reported in the main text used a partial reflector with a 40 nm film thickness with an optical density of 2.5.  We chose this attenuation of the specular reference beam by first maximizing the laser power, and then for a 500 Hz frame-rate, selecting the partial reflector that maximized the image brightness, short of saturation. The optical density of the partial reflector scales linearly with the thickness of the mask:

$$ OD = (0.0627 nm^{-1}) * \ell_{\rm mask} $$

\noindent where OD is the optical density and $\ell_{\rm mask}$ refers to the thickness of the reflecting film.

\section*{Ratiometric versus subtractive signal processing for background suppression }

Differential processing works much in the same vein as ratiometric imaging. Both techniques, differential and ratiometric, effectively isolate the areas in the field of view that change with time, and suppress the pattern associated with those that do not.  These approaches differ in that a differential image fluctuates about 0 and the ratiometric varies about 1. The subtraction of approximately equal numbers in differential imaging can cause numerical underflow around the value of 0. The division of approximately equal numbers helps circumvent this because the percent deviation (ratiometric) is larger than the absolute difference (differential). With infinite floating point precision, the outcome between ratiometric and differential imaging techniques would yield much the same result, however we prefer to implement ratiometric imaging for computational reasons.

\section*{Details about autofocus and its significance for iSCAT particle contrast quantitation }

The phase of the interference observed and thus the contrast of a scattering object measured depends very sensitively on the axial (focal) position of the scatterer, owing in large part to the contribution of the Guoy phase.  To quantify the scattering power of a particle by the  depth of its interference signal measured in an iSCAT image, one must precisely control the distance between the objective (OBJ) and the far surface of the coverslip (CS).  We use a focus stabilization strategy adapted from fluorescence microscopy \cite{bellve2014design} with two principal aims.   

This autofocus approach stabilizes the $z$ position of the coverslip exit (far surface) at the working distance of the objective. Coverslip thicknesses can vary as much as 5 $\mu$m.  So, a control strategy that sets the translation stage to a predefined $z$ position cannot succeed generally in placing a scatterer landing on the coverslip at the working distance of the objective.  Also important, autofocusing ensures reproducibility in the definition of the focal plane between experiments.  Electronic translation stages suffer from hysteresis error and sample drift.  The active feedback of autofocus system minimizes the influence of these effects.  

Fig.~\ref{fig:auto_focus} schematically illustrates autofocus signal afforded by the displacement of the reflected image of a laser beam on a digital camera.  The location of this spot on the camera directly measures the relative $z$-position of the coverslip-water interface.  We establish this quantity in real time by fitting a Gaussian to the vertically binned image profile.  Positioning the fit-mean quickly optimizes the focus.  An active feedback loop stabilizes it throughout the duration of the experiment with a precision of $\pm 5$ nm.  To optimize the focal position, we begin every experimental measurement by maximizing the destructive interference of an auxiliary 15nm gold nanoparticle.   

\begin{figure}[ht!]
    \centering
    \includegraphics[width=0.5\textwidth]{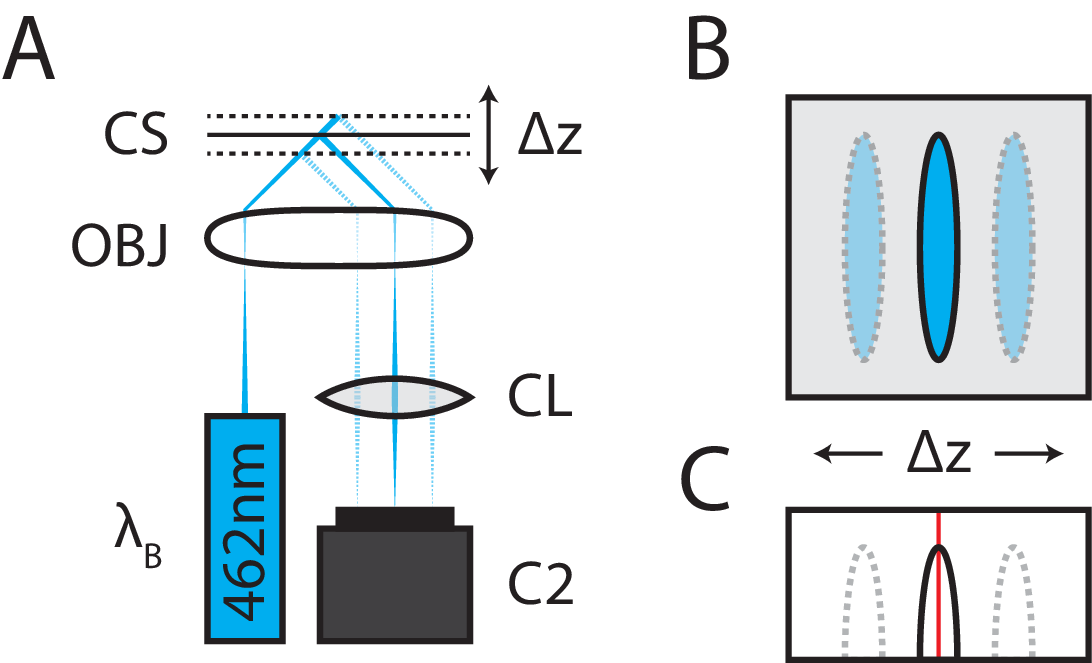}
    \caption{ Autofocus implementation schematic including: 
    (A) optical configuration including 462nm laser ($\lambda _B$), objective (obj), coverslip (CS), cylindrical lens (CL) and camera (C2),
    (B) Representative image of the autofocus feedback on the camera, and 
    (C) Vertically binned gaussian profile of the camera image in (B).}
    \label{fig:auto_focus}
\end{figure}

\section*{Autotracking TEM images, factors that contribute to the departure from a normal distribution }

We apply a Circular Hough Transform (CHT) algorithm to find dark circles in the TEM images. This algorithm sets the following conditions of measurement:
\begin{itemize}
\item Polarity of the circles must be dark
\item Sensitivity = 0.8
\item Edge threshold = 0.15
\end{itemize}

The CHT algorithm also requires upper and lower radius bounds.  When particle landings present distinctively, specifying a narrow interval of accepted radii minimizes false positives.  We specify this parameter in terms of image widths measured in CMOS pixels.  TEM images are recorded at an instrument-maximum, 200K magnification which corresponds to 0.4 nm per pixel.  Bounds specified as radii in the basis of pixels convert to particle diameter in nm based on the magnification of the TEM, ${\rm \rightarrow 2*radius(pixels)*0.4nm/pixel}$:

\begin{itemize}
\item 5nm:   [4 px, 15 px] $\rightarrow$ [3.2 nm, 12.1 nm]
\item 10nm: [8 px, 18 px] $\rightarrow$ [6.5 nm, 14.5 nm]
\item 15nm: [7 px, 35 px] $\rightarrow$ [5.7 nm, 28.2 nm]
\end{itemize}

The reliability of the CHT to recognize a Gaussian circle suffers below a lower limit of 5 pixels.  To digitally count particles in the TEM image of 5 nm gold nanoparticles, we push this limit to 4 pixels, which places a lower bound of 3.2 nm on the distribution of particle sizes measured automatically in our high-throughput particle tracking algorithm.  

The TEM manufacturer specifies a $\pm 10$\% fidelity in its image representation of a nanoparticle size.  The absolute variation thus grows with particle size, calling for a proportionally larger CHT acceptance radius for 10 and 15 nm particles.  Figure~\ref{fig:tem_supplemental} shows a sample TEM image of 15 nm gold nanoparticles as recorded by the instrument, together with an indication of the particles located and measured by our CHT algorithm.

\begin{figure}[ht!]
    \centering
    \includegraphics[width=1\textwidth]{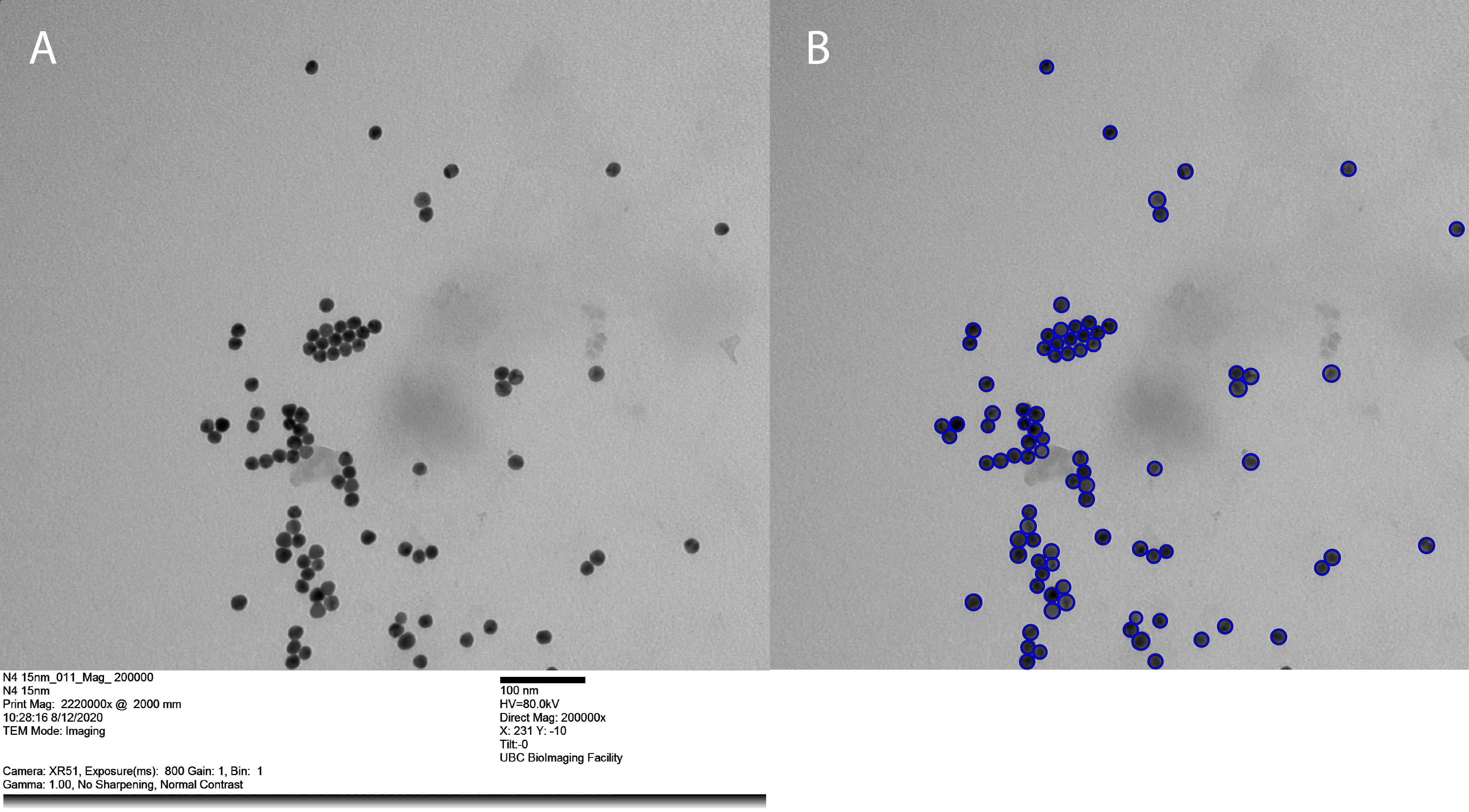}
 \caption{   (A) Raw TEM image file and (B) particles identified and measured indicated by blue circles.}
    \label{fig:tem_supplemental}
\end{figure}

\section*{Size determination accuracy of AuNPs by iSCAT}

The dynamic range of the CMOS camera and the arithmetic of binary image processing sets a limit of 1/256 on the smallest possible difference between any two individual 8 bit contrast signals collected by ratiometric processing.  This detectable difference scales linearly with the number of ratiometric frames combined to form an image.  The contrast dynamic range coverts to one of particle-diameter by reference to the constant that calibrates the contrast distributions observed by iSCAT by the distributions of directly measured TEM particle diameters ${\rm4.6*10^{-5} nm^{-3}}$, yielding an expression for measurement accuracy as a function of the size of the bins of frames combined and divided to form a ratiometric image:  $${\rm Accuracy [nm] = \sqrt[3]{\frac{\frac{1}{2^{\rm bits}*n_{\rm ratiometric}}}{4.6*10^{-5} \rm nm^{-3}}}}$$

\noindent where $2^{\rm bits}$ refers to the bit depth of the camera and $n_{\rm ratiometric}$ denotes the number of images averaged in the ratiometric stack.

The ultimate measurement accuracy depends on the dynamic range of the camera and the ratiometric bin size.   Figure \ref{fig:single_particle_accuracy} show how this accuracy varies with ratiometric bin size for standard CMOS bit depths of 8, 10 and 12.  For our 8-bit depth and ratiometric bin size of 50 images, we attain a limiting iSCAT size determination accuracy of 1.19nm.

\begin{figure}[ht!]
    \centering
    \includegraphics[width=0.8\textwidth]{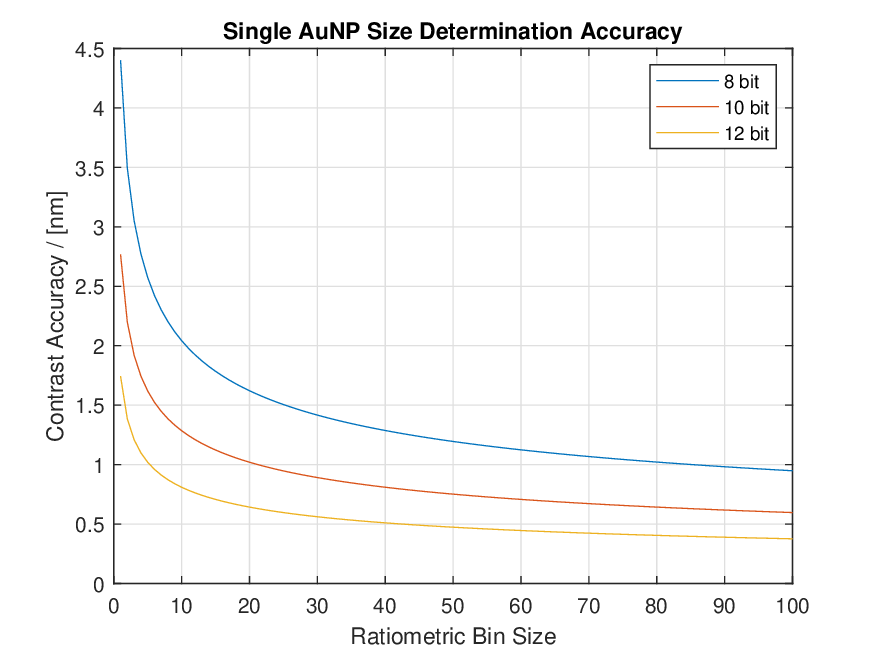}
    \caption{ Gold nanoparticle size determination accuracy as a function of camera bit depth and ratiometric binning.}
    \label{fig:single_particle_accuracy}
\end{figure}

\pagebreak

\section*{ iSCAT contrast as a mass photometry of nanoparticle distributions in solution }

Work in the Kukura group introduced iSCAT as a routine quantitative tool for the mass photometry of proteins in solution.  Calibration of the iSCAT contrast formed by particular sets of biomolecules gains precision from their well-defined molecular masses.  Yet, even the measured peak of a contrast distribution calibrated to a single protein determines the scattering mass to no better than 2\%.\cite{Young2018}   The full distributions of contrast measured in the iSCAT images of single macromolecules span substantially broader full-widths at half-maximum.  

Procedures for the synthesis of colloidal gold nanoparticles form tuned particle size distributions of varying width and average mass.\cite{WOS:000264502000001,WOS:000266615400010}  However, measures of iSCAT contrast for such solutions establishes positions of mean contrast to a precision similar to those found for mono-disperse solutions of macromolecules.  Comparing iSCAT contrasts as measures of nanoparticle volume to nanoparticle diameters observed for the same solutions by TEM, calibrates the mean contrast to a mean diameter.  The uncertainty in the mean contrast, typically less than 2\% at a 95\% confidence level maps the observed distribution of interferometric contrast for a given sample to a precise estimate of the distribution of particle sizes in that sample, broadened slightly by the same factors that vary the iSCAT contrast of particles with a single molecular mass, in other words a mass photometry of nanoparticle size distributions.  

\section*{Measurement of particle count on a surface as a gauge of concentration in solution:  Transport limitation} 

The observed rate of gold nanoparticle binding to a poly-D-lysine coated coverslip decreases much faster than deposition decreases the bulk concentration of the solution.  Such fall-off in deposition rate clearly fails to conform with a Langmuir adsorption model.  The evident size selectivity of this non-Langmuir behaviour seems to suggest a deposition process that is limited by the transport properties of nanoparticles in quiescent solutions.\cite{wo2016determining}  

Let us consider the first one-second of adsorption as it effects two successive 20 $\mu$m disks in the solution immediately above the coverslip.  We shall figure finite processes in these volume elements on this timescale as representative of differential transport associated with coverslip binding and diffusion of gold nanoparticles.  We measure initial rates for solutions of known bulk concentration, and can use the Stokes-Einstein equation to estimate gold nanoparticle diffusion coefficients \cite{midelet2017sedimentation}.   

Table \ref{tab:depletion} summarizes these diffusion coefficients, together with the initial rates observed for colloidal nanoparticle solutions of the indicated concentrations.  Here, we also figure the number of particles initially present in the two solution disks, along with the count of particles in the first disk after the depletion associated with one second of adsorption.  The decrease in nanoparticle density creates a concentration gradient on the assumed length scale of 20 $\mu$m between the second and first solvent layers.  This drives a Fick's Law diffusion of particles from the second layer to the first.  We can compare this estimate of the rate nanoparticle diffusion stimulated by initial binding with the rate of binding itself to get an idea of the net depletion in a quiescent solution under these conditions.  We see that diffusion of nanoparticles from the bulk replenishes only 12.5 percent of the 5 nm nanoparticles lost to adsorption from a layer of solvent extending 20 $\mu$m immediately above the coverslip.  Replenishment is slower for 10 and 15 nm particles, restoring only 6.5 and 3.6 percent of the particles removed by binding to the coverslip.  This size selectivity conforms very well with the time evolution of the binding distribution measured by iSCAT, and shown by Figure 8 in the main text.

\begin{figure}[ht!]
    \centering
    \includegraphics[height=0.55\textwidth]{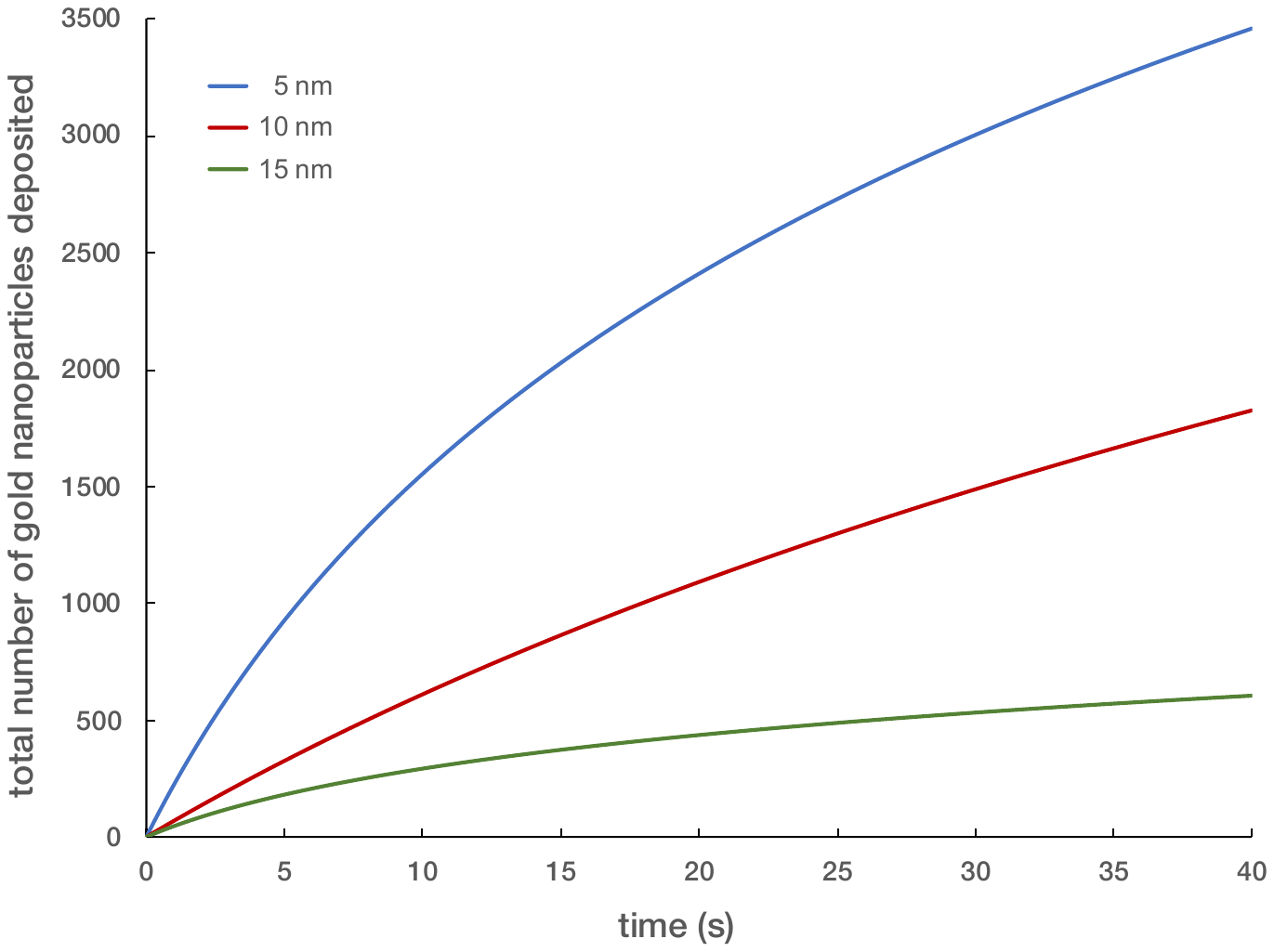}
    \caption{Cumulative numbers of 5, 10 and 15 nm gold nanoparticles deposited irreversibly on poly-D-lysine coated cover slips as a continuous function of time, from 400 $\mu$L solutions with concentrations of 1.13 nM (5 nm), 0.47 nM (10 nm) and 0.13 nM (15 nm).  
    }
    \label{fig:instrument}
\end{figure}

 \begin{table}[h!]
	\centering
	\small
	
	\begin{tabular}{ccccccccc}  
 \caption{\footnotesize{ Concentrations of gold nanoparticle solutions, of nominally 5, 10 and 15 nm.  Together with the parameters of volume elements extending from 0 to 20 $\mu$m and from 20 to 40 $\mu$m above the coverslip, defined as a finite representation of differential transport associated with coverslip binding and diffusion of gold nanoparticles.  $N(t)_0$ represents the number of particles in the first and second volume elements at $t=0$.  During the first 1 s, we assume that particle binding at the initial rate, $(dN/dt)_0$, observed in the FOV and extrapolated to the full the reference disk, reduces the number of particles in the first volume element to $N(t)$, while the concentration of the second volume element remains constant.  This creates a concentration gradient over the 20 $\mu$m separating the central planes of these volume elements, which drives a Fick's Law flux that partially replenishes the first layer by the fraction indicated for particles with diffusion constants, $D$.
}}
\label{tab:depletion}
Diam.   &   [AuNP]$_0$   &   $N(t)_0$          &   $(dN/dt)_0$   &  $(dN/dt)_0$  &   $N(t)$   &   $D$                             &   Flux       &   Fract. \\
     (nm)     &        (nM)          &     ($t=0$s)   &  (s$^{-1}$)      & (s$^{-1}$)  &   ($t=1$s)  & ($\mu$m$^2$s$^{-1}$) & (s$^{-1}$)   &     Replen\\
\toprule
5   &   1.13   &   2.69E+09   &   190   &   6.38E+08   &   2.06E+09   &   50   &   7.97E+07   &   0.125\\
10   &   0.47   &   1.12E+09   &   70   &   2.35E+08   &   8.86E+08   &   26   &   1.53E+07   &   0.065\\
15   &   0.13   &   3.10E+08   &   41   &   1.17E+08   &   1.93E+08   &   17   &   4.99E+06   &   0.036\\
	\bottomrule
	\normalsize
	\end{tabular}
\end{table}

\section*{Video micrographs}

\noindent \faGithub~{\footnotesize \url{https://github.com/grantlab-ubc/AuNP-Supplementals/blob/main/raw_movie_10nm_AuNP.mp4}}
\begin{figure}[ht!]
    \centering
    \caption{This video presents the raw video acquired by the iSCAT microscope over the course of a 10nm gold nanoparticle deposition. The experiment begins with a clean coverslip and 10 nm gold nanoparticles in solution. As the experiment proceeds, particles deposit irreversibly onto the coverslip from solution. The change in the morphology of the coverslip roughness can be seen by scrubbing through the video from start to end. The 130 second video is recorded at a full framerate of 500 Hz but visualized at 30 Hz for a net slowdown of 17x (36 minutes and 12 seconds). The field of view is $9.12 \times 9.12$ $\mu$m$^2$ with a contrast ranging from 0 to 255.   
    }
    \label{fig:movie1}
\end{figure}

\noindent \faGithub~{\footnotesize \url{https://github.com/grantlab-ubc/AuNP-Supplementals/blob/main/ratiometric_10nm_AuNP.mp4}}
\begin{figure}[ht!]
    \centering
    \caption{This short video shows the results of ratiometric processing applied to a part of video S2. Here 10 nm gold nanoparticles are clearly visible as they land and stick on the coverslip reaching a maximum contrast, then disappearing into the background. This ratiometric video is sub-sampled at the full frame rate of 500 Hz to 30 Hz. This is how the experiment appears in real-time using the operating software of the microscope. The field of view is $9.12 \times 9.12$ $\mu$m$^2$ with a contrast ranging from 0.98 to 1.01.
    }
    \label{fig:movie2}
\end{figure}

\noindent \faGithub~{\footnotesize \url{https://github.com/grantlab-ubc/AuNP-Supplementals/blob/main/particle_tracking_10nm_AuNP.mp4}}
\begin{figure}[ht!]
    \centering
    \caption{ This short video illustrates the application of our single particle tracking algorithm to part of video S2.  The algorithm identifies a total of 45 particles within the video time-frame, ranging from index 1696 to 1741.  Each particle identified is labelled with a particle database ID number for clarity. A number of particles were not identified in this video because they do not meet the strict particle tracking criteria detailed in the methods section.  Rejection criteria include a 2D Gaussian fit that is too eccentric, a poor linear fit to find the minimum particle contrast value in the contrast vs time plot ($r^2 < 0.98$), or the particle is too close to the edge of the image.1.3 seconds of video is recorded at a full frame rate of 500 Hz but visualized at 30 Hz for a net slowdown of 17x (22 seconds). The field of view is $9.12 \times 9.12$ $\mu$m$^2$ with a contrast ranging from 0.98 to 1.01.
    }
    \label{fig:movie3}
\end{figure}




\bibliography{aunp_ref_all.bib}

\end{document}